\documentclass[aps,prb,twocolumn,groupedaddress]{revtex4}
\usepackage[latin1]{inputenc}
\usepackage[dvips,final]{epsfig}
\usepackage{graphicx,bm,amsmath}

\begin{document}


\title{Omnidirectional broadband insulating device for flexural waves in thin plates}
\author{Alfonso Climente}
\author{Daniel Torrent}
\author{Jos\'e S\'anchez-Dehesa}
\affiliation{Wave Phenomena Group, Universitat Polit\`{e}cnica de Val\`{e}ncia, C/Camino de vera s.n. (edificio 7F), ES-46022 Valencia, Spain}

\date{\today}

\begin{abstract}
This work presents a gradient index device for insulating from vibrations a circular area of a thin plate. The gradient of the refractive index is achieved by locally
changing the thickness of the plate, exploiting in this way the thickness-dependence of the dispersion relation of flexural waves in thin plates. A well-like thickness
profile in an annular region of the plate is used to mimic the combination of an attractive and repulsive potentials, focusing waves at its bottom and dissipating them by means of a properly designed absorptive layer placed on top of the plate. The central area is therefore isolated from vibrations while they are dissipated at the
bottom of the well. Simulations have been done using the multilayer multiple scattering method and the results prove their broadband efficiency and omnidirectional properties. 
\end{abstract}

\pacs{}

\maketitle


\section{Introduction}
The study of the propagation of flexural waves in thin plates has received an increasing attention recently, due to the possibilities in their control offered by new structures like phononic crystals \cite{hsu2006efficient,mcphedran2009platonic,Farhat2010,pierre2010negative,lenteTTWu,Farhat2010a}, arrays of attached resonators \cite{oudich2010sonic,Xiao2012,Torrent2013Graphene} or transformation-coordinate based devices \cite{Farhat2009,Farhat2009a}. Therefore, applications such as cloaking shells \cite{Stenger2012} or negative refractive lenses \cite{pierre2010negative,bramhavar2011negative} have been experimentally demonstrated, opening the door to new and exciting devices for the control of vibrations.

In addition to all these phenomena related with classical waves in general, the propagation of flexural waves in thin plates presents the peculiarity that it can be 
controlled by means of thickness variations, as was theoretically studied by V.V. Krylov \cite{krylov2004theory}. Known as wedges, these structures produce a gradual reduction in the velocity of the wave by changing the plate's local thickness. Experimental investigations have been carried out on a variety of plate-like and beam-like structures \cite{krylov2007test,oboy2010test2}. 

The principle of wave control by locally changing the plate's thickness has been applied to achieve efficient damping of flexural waves in plate-like structures using the so called ``acoustic black holes'' (ABH) \cite{georgiev2011elliptical}, which consist of specifically designed pits with small pieces of absorbing materials attached in the middle \cite{ross1959damping}. D.J. O'Boy introduced then a modification in the ABH by perforating a hole in the center of the pit \cite{o2011tapered} and E.P. Bowyer studied the effect of imperfections in the fabrication method of the ABH \cite{bowyer2012imperfections} and the result of placing multiple ABH together in a plate \cite{bowyer2013multipletaperedbh}. For more information on these topics, see Ref. \onlinecite{krylov2012resumen}.

Based on the work introduced by V.V. Krylov et al. \cite{krylov2004theory,krylov2007test,oboy2010test2,georgiev2011elliptical} and using the Ross-Ungar-Kerwin (RKU) model \cite{ross1959damping}, in this work a new device is designed and numerically tested. The objective of the proposed device is to isolate a given area in a thin plate from vibrations. This objective is accomplished by surrounding it by a properly designed thickness-inhomogeneous region which will attract and dissipate vibrations on the plate, accomplishing then a double objective. From one side, the central region will be properly isolated, from the other one, the device will dissipate the vibrations on the plate. 

The paper is organized as follows. Section \ref{sec:theory} introduces the Kirchoff-Love approximation for modeling the behavior of flexural waves in thin plates and the method to design the structure is explained. Section \ref{sec:absorbing} explains the Ross-Kerwin-Ungar (RKU) theory used to model the behavior of an absorbing layer placed on top of an elastic plate. Following, in section \ref{sec:design} the design and optimization of the device are explained and in section \ref{sec:results} the performance of the device is tested using a numerical simulator based on a multilayer algorithm. Finally, the conclusions are presented in the last section and the multilayer algorithm is explained in Appendix A.

\section{Theory}
\label{sec:theory}

The equation of motion describing flexural waves in thin plates is modeled using the Kirchoff-Love approximation \cite{Timoshenko,Graff,Leissa}, in which the vertical displacement $W(x,y)$ of the plate is obtained from the bi-Helmholtz equation (assuming harmonic time dependence of frequency $\omega$)

		\begin{equation}\label{BiHarm}
			D \nabla^4 W(x,y)-\rho h \omega^2 W(x,y)=0
		\end{equation}
		
\noindent being $D=E h^{3}/12(1-\nu^{2})$ the flexural rigidity, $\rho$ the mass density, $h$ the thickness of the plate, $E$ the Young Modulus and $\nu$ the Poisson ratio. For plane wave propagation with wavenumber $k$ the above equation gives a quadratic dispersion relation
		
		\begin{equation}\label{wavenumber}
			k^{4} = \frac{\rho h \omega^{2}}{D},
		\end{equation}

\noindent and a phase velocity
		\begin{equation}\label{speed}
			c^{4} = \left ( \frac{\omega}{k} \right ) ^4 = \omega^2 \frac{Eh^3}{12(1-\nu^2)\rho h}.
		\end{equation}

It can be seen that the phase velocity $c$ is a function not only of the physical properties of the plate, but it also depends on its thickness. This dependence allows 
the design of gradient index devices for flexural waves easily by means of local variation of the plate's thickness. Effectively, the refractive index as a function of the
plate's thickness is

	  \begin{equation}\label{RefIndx}
			n(r,\theta) = \frac{c_b}{c(r,\theta)} = \sqrt\frac{h_{b}}{h(r,\theta)},
		\end{equation}

\noindent $c_b$ being the wave speed in the background, $h_b$ the background's thickness and $h(r,\theta)$ the position-dependent thickness. It has been assumed that all the other elastic properties of the plate remain unchanged. Equation \ref{RefIndx} describes how the refractive index increases with decreasing thickness. It is important to notice that, despite being an intrinsically dispersive medium, the refractive index does not depend on the frequency of the wave, thus the functionality of the designed devices will be limited by the accuracy of the flexural wave model only.
	
	\begin{figure}
				\includegraphics[height=40mm,width=\columnwidth]{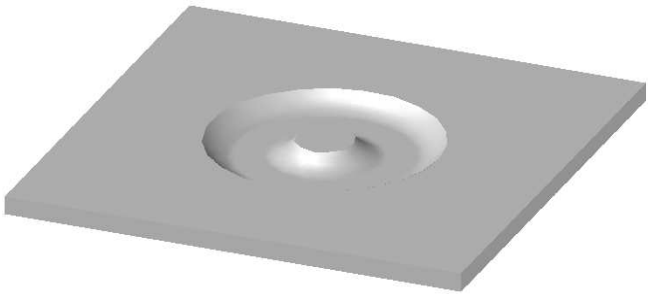}
				\caption{\label{Fig_Scheme3D}
				Schematic view of the structure studied in the present work. The central circular region is surrounded by a thickness-varying shell so that it is 
				isolated from the propagation of flexural waves on the plate. 
				}
	\end{figure}

Figure \ref{Fig_Scheme3D} shows a schematic view of the gradient index device analyzed in the present work. It consist of a circularly symmetric region in which
the thickness of the plate is gradually changed according to the desired functionality. In this case, a well-like profile is placed surrounding a central area, which
corresponds to the region that has to be isolated from external vibrations. Following the analogy with acoustic and electromagnetic waves for similar devices \cite{Narimanov, BHClimente}, the objective of the decreasing height (increasing 
refractive index) region is to act as an ``attractive'' potential, so that it tends to concentrate vibrations, while the inner region of increasing height (decreasing refractive index) will act as a ``repulsive'' potential, isolating in this way the central region. All waves traveling around this device will be concentrated at the bottom of the well where, as will be explained below, they will be dissipated. 

The proposed structure will be theoretically studied by means of a multiple scattering method, where the variation of the height is discretized and then the structure
is modeled as a multilayered shell. This method requires the application of the proper boundary conditions at each layer, for which it is necessary to know their 
physical properties. As has been previously said, the only parameter that changes from layer to layer is the plate's thickness, except at the bottom of the well, where
a dissipative material will be placed. Next section will describe the model to be used to analyze this absorbing layer.


\section{Absorbing Layer Model}
\label{sec:absorbing}
The objective of the device analyzed in this work is to concentrate waves in a given region and then dissipate them. This dissipation will be made by placing an
absorptive material of thickness $\delta$ in contact with a plate of thickness $h_a$ on the region of interest, as shown in Figure \ref{Fig_Schemelayer}. To properly 
apply to this system the multiple scattering method described in the Appendix, this region must be modeled as a single layer with a given Young modulus $E_c$, 
thickness $h_c$, Poisson ration $\nu_c$ and density $\rho_c$. Absorption is introduced in the model by adding a complex part in the Young modulus called the
loss factor $\eta$, such that $\hat{E}=E(1+i\eta)$, therefore the model must also provide the composite loss factor $\eta_c$.
	
		\begin{figure}
				\includegraphics[height=20mm,width=65mm]{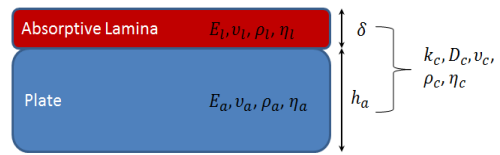}
				\caption{\label{Fig_Schemelayer} Geometry employed to dissipate the vibrations on the plate. An absorbing layer (thickness $\delta$) is placed on the top of the plate (thickness $h_a$). Each layer has its own elastic parameters that combine to produce a composite material (thickness $h_a$), with new elastic parameters. This effective material is the one considered in the model.}
		\end{figure}
		
Using the RKU model \cite{ross1959damping} it is possible describe the system plate-absorving material as a single composite. This model states that the wavenumber of the composite material is

		\begin{equation}\label{layer_kc}
			k_c^4=\frac{12\omega^2\rho_a(1-\nu_a^2)}{E_a h_a^2}\left[\frac{1+\rho_r h_r}{(1-i\eta_a)+(1-i\eta_l) h_r E_r \alpha}\right]
		\end{equation}

\noindent where the subindices $a$ and $l$ stands for the parameters of the plate and absorbing layer, respectively, and $\rho_r=\rho_l/\rho_a$, $h_r=\delta/h_a$, $E_r=E_l/E_a$ and $\alpha=3+6h_r+4h_r^2$. 

The thickness of the composite is simply the total thickness of the two layers
		
		\begin{equation}\label{layer_hc}
			h_c=h_a+\delta
		\end{equation}
while its density is the volume average of the densities of the two materials
		\begin{equation}\label{layer_rhoc}
			\rho_c=\frac{\rho_ah_a+\rho_l\delta}{h_c}.
		\end{equation}
Finally, it is assumed that the Poisson ration of the composite is the same as that of the plate, thus		
		\begin{equation}\label{layer_nuc}
			\nu_c=\nu_a
		\end{equation}
which completely characterizes the composite. From Eqs.(\ref{layer_kc})-(\ref{layer_nuc}) and (\ref{wavenumber}), the flexural rigidity is obtained as	
		
		\begin{equation}\label{layer_Dc}
		\begin{array}{l l}
			D_c & =\frac{\omega^2\rho_c h_c}{k_c^4} \\ 
			    & =\frac{h_a^3}{12(1-\nu_a^2)}E_a\left[(1+h_r E_r\alpha)-i(\eta_a+\eta_l h_r E_r \alpha)\right]
		\end{array}
		\end{equation}
		
Finally, knowing that $E_c=D_c12(1-\nu_c^2)/h_c^3$, the Young modulus of the composite is obtained and its loss factor is found as
		
	  \begin{equation}\label{layer_etac}
			\eta_c=\frac{\Im\left\{E_c\right\}}{\Re\left\{E_c\right\}}=\frac{\eta_a+\eta_l h_r E_r \alpha}{1+h_rE_r\alpha}
		\end{equation}
		
\begin{figure}
				\includegraphics[height=65mm,width=75mm]{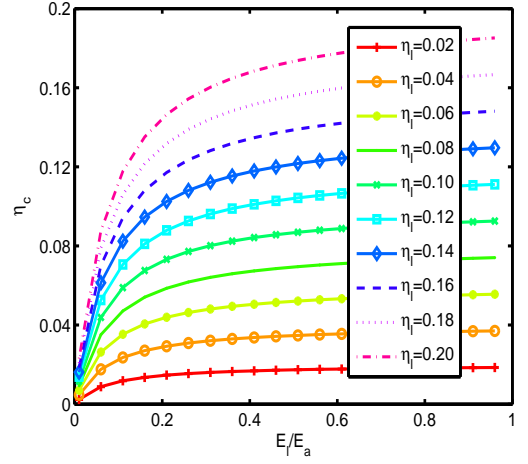}
				\caption{\label{Fig_etaC1}(Color Online)
				Variation of the composite loss factor $\eta_c$ with the normalized Young Modulus $E_r=E_l/E_a$, for different values of the loss factor of the absorptive layer $\eta_l$ and for $h_a=0.5mm$ and $\delta=0.5mm$.
				}
	\end{figure}

In this work, an aluminum plate and a molded Polystyrene layer has been used. Table \ref{Tab_Mat} shows the materials' properties \cite{MatVictor}. Figure \ref{Fig_etaC1} shows how the variation of the composite loss factor with the relative Young modulus $E_r=E_l/E_a$ for different values of $\eta_l$. The thickness of the aluminum plate is $h_a=0.5mm$ and the thickness of the layer is $\delta=0.5mm$. As can be seen, the composite loss factor $\eta_c$ approaches the loss factor
of the absorptive layer $\eta_l$ as the normalized Young Modulus $E_r>> 0.5$, as expected from Eq (\ref{layer_etac}). 

\begin{figure}
				\includegraphics[height=65mm,width=75mm]{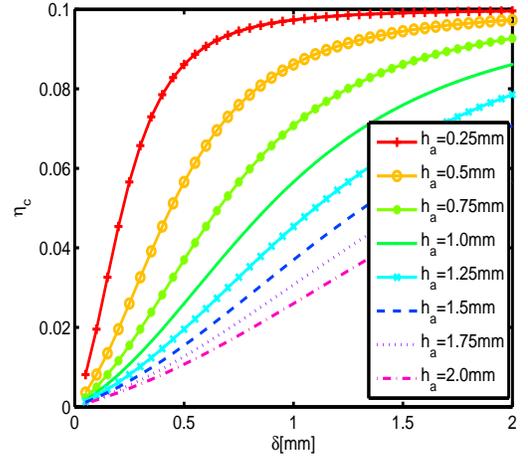}
				\caption{\label{Fig_etaC2}(Color Online)
				Variation of the composite loss factor $\eta_c$ with the thickness of the layer $\delta$ for different plate thickness $h_a$.
				}
	\end{figure}

Figure \ref{Fig_etaC2} shows how the loss factor changes with the thickness of the plate and the layer. In this case, the material properties are the ones from Table \ref{Tab_Mat}. It is seen that when the thickness of the plate decreases, a minor change in the thickness of the layer changes the loss factor greatly. Given that the thickness of both the absorptive layer
and the plate are the properties that can be better tailored, they will be optimized as explained in next section to maximize the efficiency of the device.

\begin{table}
			\caption{\label{Tab_Mat}
				Elastic Parameters of the materials used in this work
			}
			\begin{tabular}{|l| r r|}
			\hline
			              & Aluminum 			& Molded Polyester 	\\ \hline\hline
				$E$        	& 78.97Gpa 			& 7.8Gp 		 				\\ \hline
			  $\rho$     	& 2700 $kg/m^3$	& 1400 $kg/m^3$			\\ \hline
			  $\nu$ 	 		& 0.33					& 0.34							\\ \hline
			  $\eta$	 		& 0.0001				& 0.1					      \\ 
			\hline
			\end{tabular}
	\end{table}

\section{Design and Optimization}					
\label{sec:design}		
\begin{figure}
				\includegraphics[height=80mm,width=50mm]{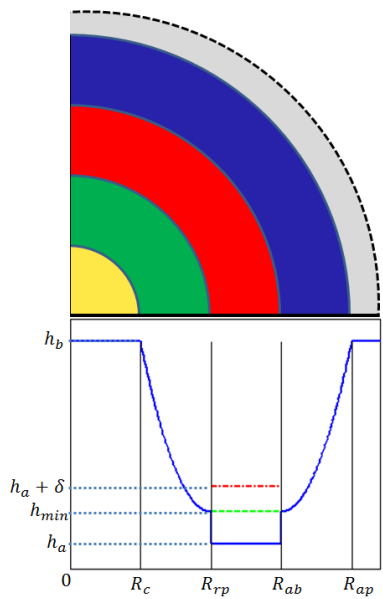}
				\caption{\label{Fig_Scheme2D}(Color Online)
				Different regions defined in the gradient index device (a) and variation of the thickness of the plate (b). The core is defined by $r<R_{c}$ (yellow) and corresponds to the area to be isolated from vibrations. Region $R_{c}<r\leq R_{rp}$ is the repulsive potential shell (green), region $R_{rp}<r\leq R_{ab}$ is the absorbing shell (red) and region $R_{ab}<r\leq R_{rp}$ is the attractive potential shell (blue). The grey region corresponds to the background and it extents towards infinity. 
				}
	\end{figure}

The full analyzed device consists then of five different regions as shown in Figure \ref{Fig_Scheme2D}a, where each region is drawn in a different color. Figure \ref{Fig_Scheme2D}b shows the variation of the thickness of the plate according to the following function

	\begin{equation}\label{ThickVsRadius}
			h(r)=\left\{\begin{array}{l c}
			h=h_b                                                           &        r \leq R_{c}\\ 
			h=\frac{h_{b}-h_{min}}{(R_{c }-R_{rp})^2}(r-R_{rp})^2+h_{min}   & R_{c} <r \leq R_{rp}\\ 
			h=h_{min}                                                       & R_{rp}<r \leq R_{ab}\\ 
			h=\frac{h_{b}-h_{min}}{(R_{ap}-R_{ab})^2}(r-R_{ab})^2+h_{min}   & R_{ab}<r \leq R_{ap}\\ 
			h=h_b                                                           & R_{ap}<r
			\end{array}\right.
	\end{equation}

\noindent where $h_b=10mm$ is the thickness of the plate in the background, $h_{min}=0.5mm$ is the minimum thickness of the plate (prior to the optimization process), $R_{c}=15cm$ is the radius of the core, $R_{rp}=30cm$ is the radius of the ``repulsive potential'' shell, $R_{ab}=45cm$ is the radius of the absorptive shell and $R_{ap}=60cm$ is the radius of the ``attractive potential'' shell. 
		
The elastic properties of the materials used in this work are stated in Table \ref{Tab_Mat}. To maximize the energy transfer through all the layers, the system requires a good matching between each interface, so that the scattering of incoming waves be minimum. Boundary conditions of the Kirchoff-Love approximation are continuity of the displacement $W$, its radial derivative $\partial_rW$, the conservation of the bending moment and the generalized Kirchoff stress, as given by 
Eq. (\ref{BoundaryCond_W}-\ref{BoundaryCond_Vr}) in Appendix A. These equations are functions of the flexural rigidity $D(r)$ and the wavenumber $k(r)$ ($\nu$ does not change), so that, in order to minimize reflections when changing from one region to the other one, these values have to be continuous. 

Although the thickness values given from Eq. (\ref{ThickVsRadius}) provides this continuity, once the effect of the absorbing layer is added, a mismatch between the layers surrounding the absorptive shell occurs. An optimization process solves this problem.  Two parameters have been tailored, the thickness of the absorbing layer ($\delta$) and the thickness of the absorbing plate ($h_a\leq h_{min}$). Figure \ref{Fig_Scheme2D}b shows, in the absorption region, the original value $h_{min}$ (green dashed curve), the new thickness value $h_a<h_{min}$ (blue continuous curve) and, finally, and the total thickness after adding the absorbing layer $h_a+\delta$ (red dash-dotted curve). The goal was to obtain a relative error for the flexural rigidity and the wavenumber
		
		\begin{eqnarray}\label{Errors}
			\epsilon_{D(r)}=|D(r)_{target}-D(r)|/D(r)_{target} 	\\
			\epsilon_{k(r)}=|k(r)_{target}-k(r)|/k(r)_{target} 
		\end{eqnarray}

\noindent of less than $0.1\%$. After the optimization, the values for the new thickness of the plate and the layer are $h_a=0.44h_{min}$ and $\delta=0.55mm$, respectively. Also, the composite loss factor achieved was $\eta_c=0.1$.

\section{Numerical Results}	
\label{sec:results}
Using the design described in the previous section and the numerical multilayer algorithm described in Appendix A, simulations have been done to test the performance of the device. Each region has been discretized with $N=100$ layers, which has been shown to be a good approximation to the ideal continuous device for the wavelengths of interest. 

\begin{figure}
				\includegraphics[height=60mm,width=70mm]{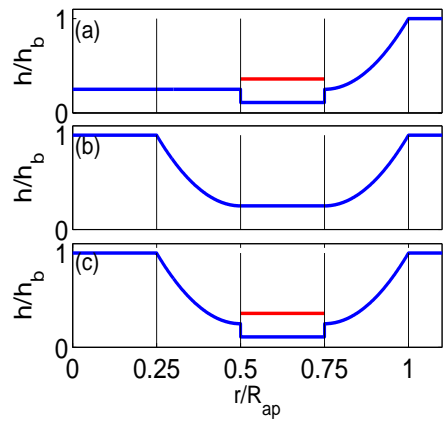}
				\caption{\label{Fif_SchemeDev} Thickness variation as a function of the distance $r$ for the three different structures tested in the present work. (a) Attractive potential with the absorptive layer. (b) Attractive and repulsive potentials without the absorptive layer. (c) Full isolating device.
	}
	\end{figure}

For comparison purposes, the three configurations shown in figure \ref{Fif_SchemeDev} have been tested. The first one is a device with the same geometrical characteristics as the designed one, but without the repulsive potential (Fig.\ref{Fif_SchemeDev}a); the second one is a device with the same geometrical characteristics as the designed one, but without the absorptive layer and the optimization (Fig.\ref{Fif_SchemeDev}b); and the third one is the designed device (Fig.\ref{Fif_SchemeDev}c).
	
\begin{figure}
				\includegraphics[height=110mm,width=80mm]{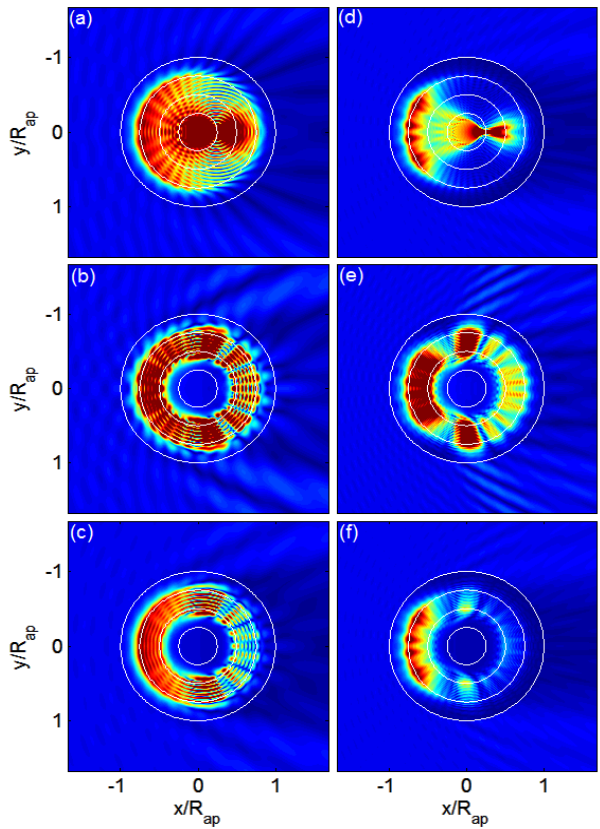}
				\caption{\label{Fig_WaveMod}(Color Online)
				Modulus of the displacement in the $z$-direction when a plane wave with wavenumber $kR_{ap}=15$ (left panels) and $kR_{ap}=35$ (right panels) impinges on the three devices tested in this work. Displacement produced by the device without the repulsive potential (a), by the device without absorption (b) and by the designed device (c).
				}
	\end{figure}

To further understand what is occurring inside the three configurations, let us consider the displacement field illustrated in figure \ref{Fig_WaveMod}. It shows the modulus of the displacement in the $z$-direction when a plane wave with wavenumbers $kR_{ap}=15$ (left panels) and $kR_{ap}=35$ (right panels), impinges on each device tested in this work. The white circles represent the boundaries defined by equation \eqref{ThickVsRadius}. Panels (a) and (d) represent the plot of the device without the repulsive potential. Notice that, although the system presents absorption, the wave is focused into the core and a high amplitude is achieved. Panels (b) and (e), are for the device without the absorptive layer, but adding the repulsive potential shell. It is seen that the wave is expelled from the core and the amplitude decreases in comparison to the previous panels, even though it does not have absorption. Finally, panels (c) and (f), correspond to the designed device. By introducing the absorptive layer, the wave amplitude in the core is further reduced. Notice that the panel in the second row has the same pattern as the one in the third row but without the attenuation.
	
As a parameter used to compare results, let us introduce the vibration average in the region $i$ defined as 

	\begin{equation}\label{VibrationAvg}
			<|W|^2>_{i}=\frac{1}{S_c}\iint_{S_c}\left | W \right(r,\theta) |^2dS
	\end{equation}
	
\noindent where $S_i$ is the area defined by $R_{i}<r<R_{i+1}$, with $R_{i} \in [\infty, R_{ap},R_{ab},R_{rp},R_{c},0]$. 

%
%
%
%

\begin{figure}
				\includegraphics[height=60mm,width=70mm]{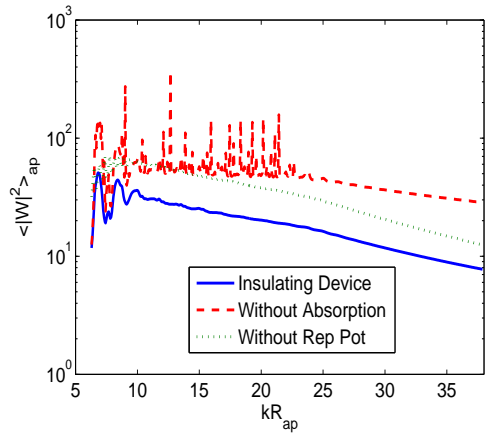}
				\caption{\label{Fig_VibAvgAP}(Color Online)
				Vibration average in the attractive potential shell for the designed device (blue curve), the device without absorption (red dashed curve) and the device without repulsive potential shell (green dotted curve).
				}
	\end{figure}

\begin{figure}
				\includegraphics[height=60mm,width=70mm]{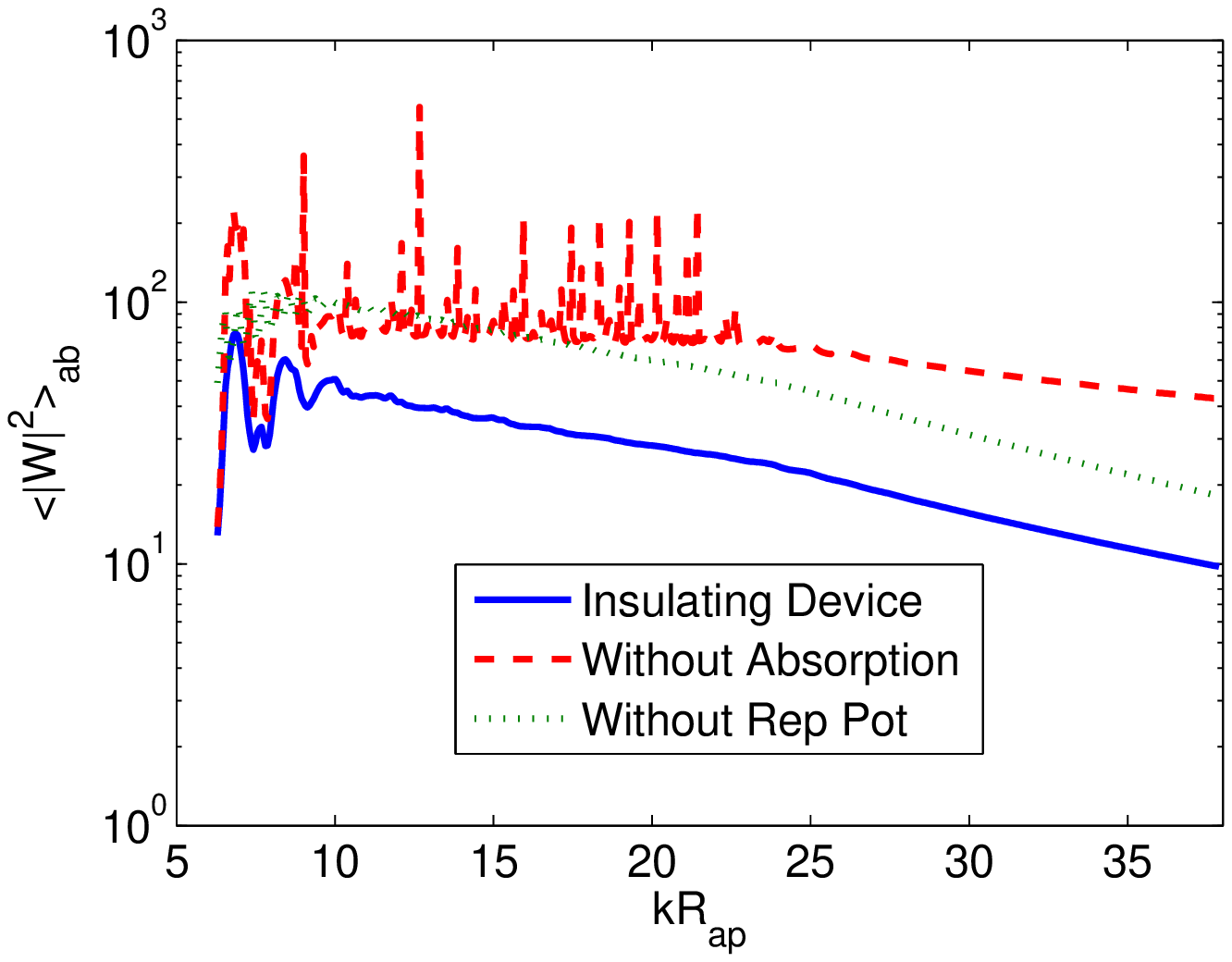}
				\caption{\label{Fig_VibAvgAB}(Color Online)
				Vibration average in the absorptive shell for the designed device (blue curve), the device without absorption (red dashed curve) and the device without repulsive potential shell (green dotted curve).
				}
	\end{figure}

\begin{figure}
				\includegraphics[height=60mm,width=70mm]{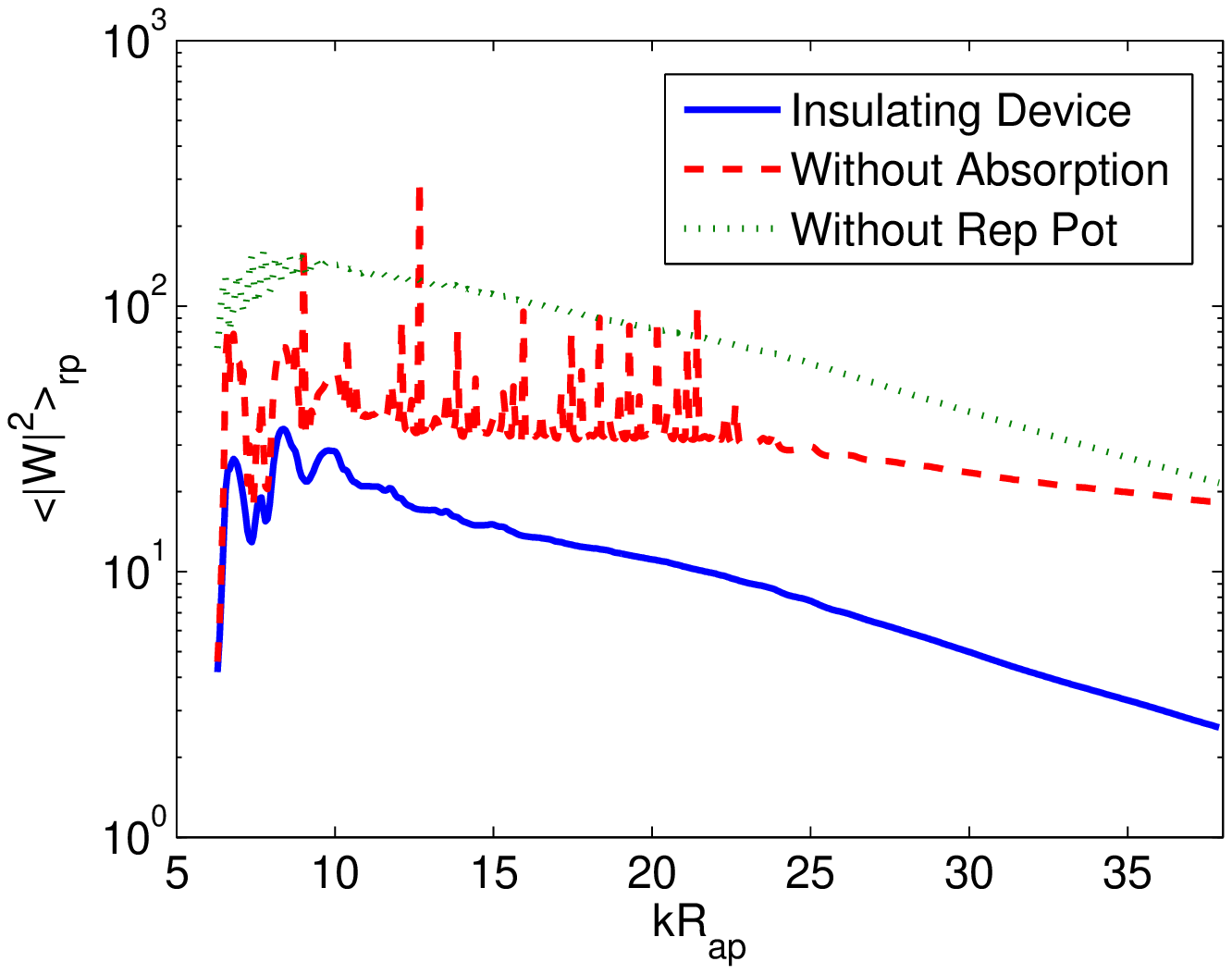}
				\caption{\label{Fig_VibAvgRP}(Color Online)
				Vibration average in the repulsive shell for the designed device (blue curve), the device without absorption (red dashed curve) and the device without repulsive potential shell (green dotted curve).
				}
	\end{figure}

\begin{figure}
				\includegraphics[height=60mm,width=70mm]{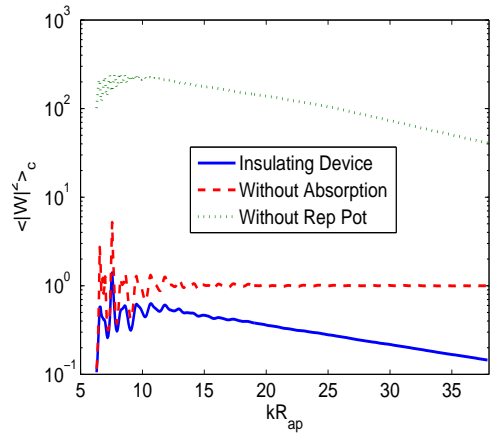}
				\caption{\label{Fig_VibAvgC}(Color Online)
				Vibration average in the core for the designed device (blue curve), the device without absorption (red dashed curve) and the device without repulsive potential shell (green dotted curve).
				}
	\end{figure}

Figure \ref{Fig_VibAvgAP},\ref{Fig_VibAvgAB}, \ref{Fig_VibAvgRP} and \ref{Fig_VibAvgC} show the vibration average produced in the four regions defining the device: the attractive potential shell, the absorbing shell, the repulsive shell and the core, respectively. Each figure shows the values of the vibration average for the designed device (blue curve), the device without absorption shell (red curve) and the device without the repulsive potential shell (green curve). Notice that in general, the vibration average decreases with increasing frequency. This is due to the frequency response of the absorptive layer. 

Figures \ref{Fig_VibAvgAP} and \ref{Fig_VibAvgAB} show that the values of the configuration without absorption (red dashed curve) are higher because of the lack of absorption, and some peaks due to resonances of the structure can also be observed. The configuration without the repulsive potential shell (green dotted curve) presents a focusing effect as seen in figure \ref{Fig_WaveMod}, and therefore more energy is concentrated. Finally, the designed device (blue curve) has the least vibration average, due to the combined effect of the absorbing layer and the repulsive potential. 

Figure \ref{Fig_VibAvgRP} shows that the field in  the configuration without the repulsive potential shell (green dotted curve) has the most vibration average due to the focusing effect that can be seen in figure \ref{Fig_WaveMod}. On the other hand, the values for the device without absorption (red dashed curve) are greater than the ones of the designed device (blue curve) due to the combined effect of the repulsive potential and the absorption. 

Finally, figure \ref{Fig_VibAvgC} shows the efficiency of the designed device. It is seen that the two configurations with the repulsive potential shell have much lower values, produced by the change in thickness and by the focusing effect. Overall, the designed device (blue curve) has the minimum vibration average in all the regions, which shows its efficiency not only for dissipating vibration energy but also for isolating a given region from these vibrations.

\section{Summary}
\label{sec:conclusions}
In this work, the omnidirectional and broadband performance of a gradient index flexural wave device has been demonstrated. Also, an explanation of the method to control the refractive index with the thickness has been described. Moreover, during this process, the RKU model to treat absorptive lamina over elastic material has been comprehensibly studied. With these results an optimization process has been done to maximize the energy transfer between the regions and the numeric multilayer algorithm based on multiple scattering used in the simulations has been explained. Also, the characteristics of the material are existent in the nature and the simulated design can be feasible to obtain, so an experimental verification can be performed in a future time.

\section{Acknowledgments}
\begin{acknowledgments}
	This work has been supported by the U.S Office of Naval Research under Grant No. N000140910554.
\end{acknowledgments}

\appendix
\section{Simulation Method}
\label{sec:appendix}
\begin{figure}
				\includegraphics[height=30mm,width=75mm]{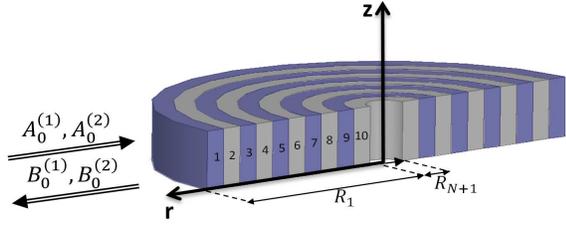}
				\caption{\label{Fig_SchemeML}(Color Online)
				Multilayered structure employed in the multiple scattering algorithm with $N=10$ layers. The background layer is $n=0$ and corresponds to the
				region $r>R_{1}$) and the core layer is $n=N+1$ and corresponds to the region $r<R_{N+1}$.
				}
	\end{figure}

The structures analyzed in the present work are radially inhomogeneous, that is, their parameters depend only on the radial coordinate. The model employed
to make numerical simulations is the multilayer scattering method, where the continuous variation of the parameters has been discretized in a number $N$ of homogeneous cylindrically symmetric layers \cite{cai2008acoustical}. Figure \ref{Fig_SchemeML} shows the cylindrical structure discretized in $N=10$ layers each one with different elastic properties. Note that the plate thickness is not a geometrical parameter anymore, but it is introduced in the boundary conditions through the elastic properties of the material. The layers are numbered such that he background corresponds to $n=0$ and the core layer corresponds to $n=N+1$.

The displacement $W_{n}$ in each layer is a solution of Eq. (\ref{BiHarm}) then, in polar coordinates $(r,\phi)$, it can be expanded as 

	  \begin{equation}\label{BiharmSol}
			W_n(r,\theta)=W_n^{(1)}(r,\theta)+W_n^{(2)}(r,\theta)
		\end{equation}

\noindent where $W_n^{(1)}(r,\theta)$ and $W_n^{(2)}(r,\theta)$ are solutions of the Helmholtz and Modified Helmholtz equations, respectively, corresponding to
Bessel and Hankel functions (and their modified versions) \cite{Leissa}. Thus, the solution can be explicitly expressed as

		\begin{equation}\label{BesselExp}
		\begin{array}{l r}
				W_{n}(r,\theta) = & \sum_{q}\left [ A_{n,q}^{(1)} J_{q}(k_{n}r)+ A_{n,q}^{(2)} I_{q}(k_{n}r) \right ] e^{iq\theta}+ \\ 
										      & \sum_{q}\left [ B_{n,q}^{(1)} H_{q}(k_{n}r)+ B_{n,q}^{(2)} K_{q}(k_{n}r) \right ] e^{iq\theta}
	  \end{array}
		\end{equation}
		
\noindent being  $A_{n,q}^{(1)}, A_{n,q}^{(2)}, B_{n,q}^{(1)}, B_{n,q}^{(2)}$ the coefficients of the expansion ($A$ for the incoming wave towards the center of the cylinder and $B$ for the scattered one) and $k_{n}$ being the wavenumber in the $n$-th layer. 

The objective now is to relate the coefficients of each layer with the ones of the previous and next layer. Defining

			\begin{equation}\label{MLal1}
				\hat{A}_{n} =
				\begin{bmatrix}
				A_{n,q}^{(1)}\\ 
				A_{n,q}^{(2)}
				\end{bmatrix}	,
				\hat{B}_{n} =
				\begin{bmatrix}
					B_{n,q}^{(1)}\\ 
					B_{n,q}^{(2)}
				\end{bmatrix} \nonumber
			\end{equation}
			
\noindent it can be seen from Figure \ref{Fig_SchemeML} that the relation between the coefficients of layers $n$ and $n-1$ is given by

		\begin{eqnarray}\label{MLalAB}
				\hat{A}_{n}   = T_{n-1n} \cdot \hat{A}_{n-1}+R_{nn-1} \cdot \hat{B}_{n} \\
				\hat{B}_{n-1} = R_{n-1n} \cdot \hat{A}_{n-1}+T_{nn-1} \cdot \hat{B}_{n}
		\end{eqnarray}

\noindent being $T_{n-1n}$ and $R_{n-1n}$ the reflection and transmission coefficient matrix (size 2x2) from  layer $n-1$ to $n$, respectively, and that will be 
computed later. Defining the layer elastic impedance matrices (size 2x2) as $\hat{B_{n}}=Z_{n} \cdot \hat{A_{n}}$, the above equations read as 

		\begin{eqnarray}\label{MLalA}
				\hat{A}_{n} = & (I-R_{nn-1} \cdot Z_{n})^{-1} \cdot T_{n-1n} 					 \cdot \hat{A}_{n-1} \\
				\hat{A}_{n} = & (T_{nn-1} \cdot Z_{n})^{-1} \cdot (Z_{n-1} - R_{n-1n}) \cdot \hat{A}_{n-1}
		\end{eqnarray}

\noindent from which we can obtain the recursive relation for the coefficient $Z_{n}$ as

		\begin{eqnarray}\label{MLalZX}
				Z_{n-1} = R_{n-1n} + T_{nn-1} \cdot Z_{n} \cdot X_{n} \\
				X_{n} = (I-R_{nn-1} \cdot Z_{n})^{-1} \cdot T_{n-1n}
		\end{eqnarray}

Starting at the last layer $n=N$, since $\hat{B}_{N+1}=0$, the impedance in the last layer $Z_{N}$ is simply

		\begin{equation}\label{MLalRec}
				\hat{B}_{N}=R_{NN+1} \cdot \hat{A}_{N} \rightarrow Z_{N}=R_{NN+1}
		\end{equation}
		
\noindent now the iterative process continues till $n=1$, so all $Z_{n}$ and $X_{n}$ matrix are obtained and applying the following relationship

		\begin{eqnarray}\label{MLalZX2}
				\hat{B}_{n}  =Z_{n} \cdot \hat{A}_{n} \\
				\hat{A}_{n+1}=X_{n} \cdot \hat{A}_{n}
		\end{eqnarray}

\noindent the incoming and scattering coefficients of each layer can be obtained as a function of $\hat{A}_0$, which is defined by the external incident field, and once the reflection and transmission coefficients of every layer in both propagation directions (to and from the center of the structure) are known. The procedure to obtain
these coefficients is detailed below.

\begin{figure}
				\includegraphics[height=35mm,width=75mm]{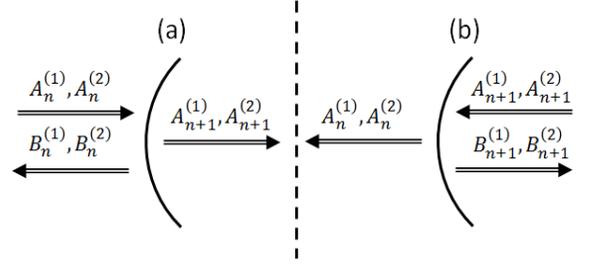}
				\caption{\label{Fig_SchemeML2}(Color Online)
				Mono-layer systems employed to obtain the reflexion and transmission matrices from layer $n$ to $n+1$ (a) and from layer $n+1$ to $n$ (b).
				}
	\end{figure}

Let us consider a single layer with only one boundary where an incoming wave $\hat{A}_{i}$ impinges on the interface. In Figure \ref{Fig_SchemeML2}a the incoming wave $\hat{A}_{n}$ travels towards the center of the cylinder, producing a reflected wave $\hat{B}_{n}$ in the opposite direction and a transmitted one $\hat{A}_{n+1}$ to the next layer. On the other hand, in Figure \ref{Fig_SchemeML2}b, the incoming wave $\hat{A}_{n+1}$ travels towards infinity, producing a reflected wave $\hat{B}_{n+1}$ towards the center of the cylinder and a transmitted one $\hat{A}_{n}$ to the previous layer.

To obtain these reflection and transmission matrices, the following boundary conditions \cite{Timoshenko,Graff,Leissa,NorrisVemula} have to be met.

		\begin{subequations}
		\begin{equation}\label{BoundaryCond_W}
				\left.W_{n}(r,\theta) \right|_{r=R_n} 														  = \left.W_{n+1}(r,\theta) \right|_{r=R_n} 		
		\end{equation}
		\begin{equation}\label{BoundaryCond_pW}
				\left.\frac{\partial W_{n}(r,\theta)}{\partial r} \right|_{r=R_n}   = \left.\frac{\partial W_{n+1}(r,\theta)}{\partial r} \right|_{r=R_n} 
		\end{equation}
		\begin{equation}\label{BoundaryCond_Mr}
				\left.M_r(W_{n}(r,\theta)) \right|_{r=R_n} 												  = \left.M_r(W_{n+1}(r,\theta)) \right|_{r=R_n} 
		\end{equation}
		\begin{equation}\label{BoundaryCond_Vr}
				\left.V_r(W_{n}(r,\theta)) \right|_{r=R_n} 												  = \left.V_r(W_{n+1}(r,\theta)) \right|_{r=R_n} 
		\end{equation}
		\end{subequations}
		
\noindent where $M_r(f)$ is the radial moment and $V_r(f)$ is the Kirchoff-Stress defined as 

		\begin{equation}\label{Mr}
			M_r(f)=-D\left[\frac{\partial^2 f}{\partial r^2}+\nu\frac{1}{r}\frac{\partial f}{\partial r} +\nu\frac{1}{r^2}\frac{\partial^2f}{\partial\theta^2}\right]
		\end{equation}

		\begin{equation}\label{Vr}
			V_r(f)=-D\frac{\partial}{\partial r}\Delta f-D(1-\nu)\frac{1}{r^2}\frac{\partial}{\partial \theta}\left(\frac{\partial^2f}{\partial{r}\partial{\theta}}-\frac{1}{r}\frac{\partial f}{\partial \theta}\right)
		\end{equation}

\noindent $D$ being the flexural rigidity and $\nu$ is the Poisson ratio. For the first system, after applying these boundary conditions, the coefficients of 
layers $n$ and $n+1$  are related by\cite{NorrisVemula}

		\begin{widetext}
		\begin{equation}\label{LayerMat1}
				\begin{bmatrix}
									H_q(\kappa_{n})			  &  					 K_q(\kappa_{n}) 			 & 						 -J_q(\kappa_{n+1}) 		& 						-I_q(\kappa_{n+1})			\\ 
				\kappa_{n}H_q^{'}(\kappa_{n}) 	&  \kappa_{n}K_q^{'}(\kappa_{n}) 	 & -\kappa_{n+1}J_q^{'}(\kappa_{n+1}) & -\kappa_{n+1}I_q^{'}(\kappa_{n+1})	\\ 
									S_{n}^{H}(\kappa_{n}) &  					 S_{n}^{K}(\kappa_{n}) & -S_{n+1}^{J}(\kappa_{n+1}) 				& -S_{n+1}^{I}(\kappa_{n+1})					\\ 
									T_{n}^{H}(\kappa_{n}) &  					 T_{n}^{K}(\kappa_{n}) & -T_{n+1}^{J}(\kappa_{n+1}) 				& -T_{n+1}^{I}(\kappa_{n+1})
				\end{bmatrix}
				\begin{Bmatrix}
				B^{(1)}_{n  ,q}\\ 
				B^{(2)}_{n  ,q}\\ 
				A^{(1)}_{n+1,q}\\ 
				A^{(2)}_{n+1,q}
				\end{Bmatrix}
				=				(-1)
				\begin{bmatrix}
				J_q(\kappa_{n}) 							& I_q(\kappa_{n})\\ 
				\kappa_{n}J_q^{'}(\kappa_{n}) & \kappa_{n}I_q^{'}(\kappa_{n})\\ 
				S_{n}^{J}(\kappa_{n}) 				& S_{n}^{I}(\kappa_{n})\\ 
				T_{n}^{J}(\kappa_{n}) 				& T_{n}^{I}(\kappa_{n})
				\end{bmatrix}
				\begin{Bmatrix}
				A^{(1)}_{n,q}\\ 
				A^{(2)}_{n,q}
				\end{Bmatrix}
		\end{equation}
		\end{widetext}

\noindent where 

		\begin{equation}\label{LayerMat_S}
		\begin{array}{l l r}
				S_{n}^{X}(\kappa_{n}) & = D_{n} &\left[  (q^2(1-\nu_{n})   \mp \kappa_{n}^2)            X_q(\kappa_{n})     \right.   \\
															&	        &\left.  -     (1-\nu_{n})                   \kappa_{n} X_q^{'}(\kappa_{n}) \right]   
		\end{array}
		\end{equation}
		
		\begin{equation}\label{LayerMat_T}
		\begin{array}{l l r}
				T_{n}^{X}(\kappa_{n}) & = D_{n} &\left[  (q^2(1-\nu_{n}) 						     )              X_q(\kappa_{n})     \right.   \\
															&	        &\left.  - (q^2(1-\nu_{n}) \pm \kappa_{n}^2) \kappa_{n} X_q^{'}(\kappa_{n}) \right] 
		\end{array}
		\end{equation}

\noindent and $X=J,I,H,K$; the upper sign is used for ($J,H$) and the lower sign for $(I,K)$ and $\kappa_{n}=k_{n}R_{n}$, $k_n$ being the wavenumber of the $n-th$ layer and $R_{n}$ the radius of the boundary between the layer $n+1$ and $n$. 

Similarly, for the second system, the coefficients of layers $n+1$ and $n$ are related by the following system of equations
	
		\begin{widetext}
		\begin{equation}\label{LayerMat2}
				\begin{bmatrix}
										J_q(\kappa_{n+1})       &  					   I_q(\kappa_{n+1}) 		     & 					 -H_q(\kappa_{n}) 	  & 					-K_q(\kappa_{n})	  	\\ 
				\kappa_{n+1}J_q^{'}(\kappa_{n+1}) 	&  \kappa_{n+1}I_q^{'}(\kappa_{n+1}) 	   & -\kappa_{n}H_q^{'}(\kappa_{n}) & -\kappa_{n}K_q^{'}(\kappa_{n})	\\ 
									S_{n+1}^{J}(\kappa_{n+1}) &  					   S_{n+1}^{I}(\kappa_{n+1}) & -S_{n}^{H}(\kappa_{n}) 				& -S_{n}^{K}(\kappa_{n})	     		\\ 
									T_{n+1}^{J}(\kappa_{n+1}) &  					   T_{n+1}^{I}(\kappa_{n+1}) & -T_{n}^{H}(\kappa_{n}) 				& -T_{n}^{K}(\kappa_{n})
				\end{bmatrix}
				\begin{Bmatrix}
				B^{(1)}_{n+1,q}\\ 
				B^{(2)}_{n+1,q}\\ 
				A^{(1)}_{n  ,q}\\ 
				A^{(2)}_{n  ,q}
				\end{Bmatrix}
				=				(-1)
				\begin{bmatrix} 
										H_q(\kappa_{n+1}) 	  		& 						K_q(\kappa_{n+1})					\\ 
				\kappa_{n+1}H_q^{'}(\kappa_{n+1}) 		& \kappa_{n+1}K_q^{'}(\kappa_{n+1})			\\ 
										S_{n+1}^{H}(\kappa_{n+1}) &    					S_{n+1}^{K}(\kappa_{n+1})	\\ 
										T_{n+1}^{H}(\kappa_{n+1}) & 						T_{n+1}^{K}(\kappa_{n+1})
				\end{bmatrix}
				\begin{Bmatrix}
				A^{(1)}_{n+1,q}\\ 
				A^{(2)}_{n+1,q}
				\end{Bmatrix}
		\end{equation}
		\end{widetext}

Now, by knowing the definition of the reflection and transmission matrices in the first system 

		\begin{eqnarray}\label{RT1}
				R_{nn+1} = \hat{B}_{n}   \cdot (\hat{A}_{n})^{-1} \\
				T_{nn+1} = \hat{A}_{n+1} \cdot (\hat{A}_{n})^{-1} 
		\end{eqnarray}
		
\noindent and the second system,

		\begin{eqnarray}\label{RT2}
				R_{n+1n} = \hat{B}_{n+1} \cdot (\hat{A}_{n+1})^{-1} \\
				T_{n+1n} = \hat{A}_{n}   \cdot (\hat{A}_{n+1})^{-1} 
		\end{eqnarray}

\noindent it is straightforward to obtain them from Eq. (\ref{LayerMat1}) and (\ref{LayerMat2}).

\end{document}